\documentclass[10pt,twocolumn,letterpaper]{article}

\usepackage{iccv}
\usepackage{times}
\usepackage{epsfig}
\usepackage{graphicx}
\usepackage{amsmath}
\usepackage{amssymb}
\usepackage{booktabs}
\usepackage{multirow}
\usepackage{bbding}
\usepackage[pagebackref=true,breaklinks=true,letterpaper=true,colorlinks,bookmarks=false]{hyperref}

\def\iccvPaperID{10773} % *** Enter the ICCV Paper ID here

\ificcvfinal\fi

\begin{document}

%%%%%%%%% TITLE
\title{Reconstructed Convolution Module Based Look-Up Tables for Efficient Image Super-Resolution}

\author{
Guandu Liu$^{1}$
\and
Yukang Ding$^{2}$\and
Mading Li$^{2}$\and
Ming Sun$^{2}$\and
Xing Wen$^{2}$\and
Bin Wang$^{1}$\footnote{Corresponding Author}
% \affiliations
\and
$^1$School of Software, Tsinghua University\\
$^2$Kuaishou\\
{\tt\small liugd21@mails.tsinghua.edu.cn}, 
{\tt\small $\{$dingyukang, limading, sunming03$\}$@kuaishou.com}\\
{\tt\small td.wenxing@gmail.com}, {\tt\small wangbins@mails.tsinghua.edu.cn}
% Guandu Liu$^{1}$\\
% % Tsinghua University\\
% % Institution1 address\\
% {\tt\small liugd21@mails.tsinghua.edu.cn}
% % For a paper whose authors are all at the same institution,
% % omit the following lines up until the closing ``}''.
% % Additional authors and addresses can be added with ``\and'',
% % just like the second author.
% % To save space, use either the email address or home page, not both
% \and
% Yukang Ding$^{2}$\\
% % Kuaishou\\
% % First line of institution2 address\\
% {\tt\small dingyukang@kuaishou.com}
}

\maketitle
% Remove page # from the first page of camera-ready.
% \ificcvfinal\thispagestyle{empty}\fi
% Giant receptive-field and tiny stored Look-UP tables for Efficient Image Super-Resolution
%%%%%%%%% ABSTRACT
\begin{abstract}
% 之前直接增大感受野，为什么造成体积增大。。已有工作忽略了哪些方面，信息交换平津。空间和深度之间的解耦。   
Look-up table(LUT)-based methods have shown the great efficacy in single image super-resolution (SR) task. 
However, previous methods ignore the essential reason of restricted receptive field (RF) size in LUT, which is caused by the interaction of space and channel features in vanilla convolution. They can only increase the RF at the cost of linearly increasing LUT size.  To enlarge RF with contained LUT sizes, we propose a novel Reconstructed Convolution(RC) module, which decouples channel-wise and spatial calculation. It can be formulated as $n^2$ 1D LUTs to maintain $n\times n$ receptive field, which is obviously smaller than $n\times n$D LUT formulated before. The LUT generated by our RC module reaches less than 1/10000 storage compared with SR-LUT baseline. The proposed Reconstructed Convolution module based LUT method, termed as RCLUT, can enlarge the RF size by 9 times than the state-of-the-art LUT-based SR method and achieve superior performance on five popular benchmark dataset. Moreover, the efficient and robust RC module can be used as a plugin to improve other LUT-based SR methods. The code is available at \url{https://github.com/liuguandu/RC-LUT}.
\end{abstract}


% figure 1 presents the trade-off of performance and storage with lut-methods
\begin{figure}[t]
\begin{center}
   \includegraphics[width=0.9\linewidth]{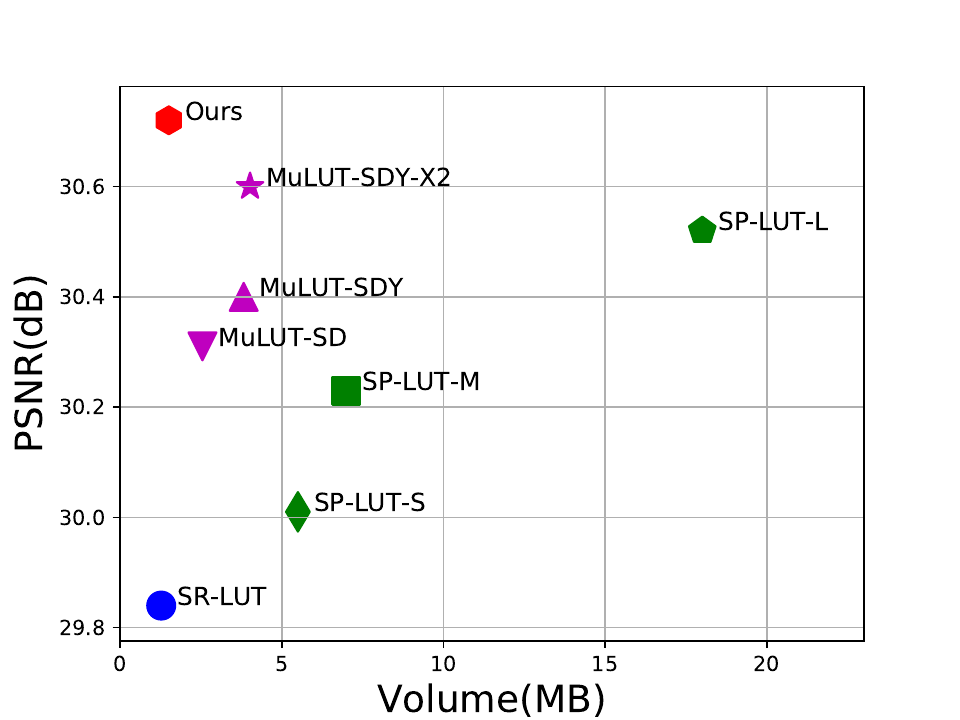}
\end{center}
   \caption{Comparison of PSNR and storage on Set5 benchmark dataset for $\times 4$ SR task. We compare our method with prior LUT-based SR methods. Our method achieves the superior performance with relatively less LUT storage.}
\label{fig:PSNR&Volume}
\end{figure}

%%%%%%%%% BODY TEXT
\section{Introduction}
\label{sec:intro}

Single image super-resolution (SR) aims to recover the high-resolution (HR) image from a low-resolution (LR) image and to bring back clearer edges, textures, and details while increasing the resolution of the input. In the past decade, deep learning-based SR methods ~\cite{RCAN, EDSR, VDSR, residual, esrgan, lai2017deep, haris2018deep, mei2020image, niu2020single} have made remarkable improvements compared with traditional SR methods (\eg, interpolation based ~\cite{bicubic}, sparse coding based ~\cite{A+, ANR, Zeyde, NE+LLE, gu2015convolutional}). However, such kind of methods often possess huge amount of parameters, which has high computation cost and cannot be practically used on devices with limited computational resources. Exploring the pratical and real-time SR solutions have been a growing trend in the single image super resolution (SISR) community.

Recently, SR methods based on look-up table (LUT) have emerging. One of the representative methods SR-LUT ~\cite{srlut21} proposes to cache the results of a trained SR network given all possible inputs to a LUT. By replacing the run-time computation of the inference process with a simpler and faster indexing operation, SR-LUT can be easily implemented on mobile devices. However, such method requires a limited receptive field (RF), as the size of the input space (\ie, the indexing capacity of the output LUT) grows exponentially with the increase of the input pixels. To be more specific, the input size of SR-LUT is $2\times 2$ (\ie, a 4D-LUT), which results in a $3\times 3$ RF by rotation ensemble, and needs $b^{2\times 2}\times r^2$ bytes ($b$ is the number of possible values of one pixel) to store its LUT in order to perform an upscale with the factor of $r$. For example, a $3\times3$ vanilla convolution will generate $255^9$ input-output mappings, which takes up 1.72 TB in LUT form.

Larger RF encourages model to capture more details of semantics and structures in single image, and it plays an essential role in training process. As for SR-LUT, the exponential growth of LUT size limits the improvement of RF significantly. To overcome this problem, MuLUT ~\cite{mulut22} and SPLUT ~\cite{splut22} both propose to cascade multi-parallel LUTs and expand the RF to $9\times 9$ and $6\times 6$, respectively. These methods try to divide the whole LUT to many sub-LUTs with linearly increasing LUT size. However, they don't delve into the most vital reason for Lut-based methods' restricted RF size. Furthermore, the RF size of these LUT-based methods still have a great distance of DNN SR methods, resulting their performances are inferior than simple DNN models, \ie FSRCNN~\cite{FSRCNN}.

As vanilla convolution fuses features in both spatial and channel dimensions, the LUT format has to traverse all possible cases of input pixels by permutation and combination. In other words, spatial depend convolution will refer to other feature points in the space neighborhood to generate input-output mapping of the current feature point.
% input-output mapping obtained by anchor feature point through spatial dependent convolution is related to other feature points in the spatial neighborhood.
Since the most critical factor limiting the performance of LUT-based methods is the spatial dependent convolution, we consider that why not decouple the spatial and channel calculations of convolution process and store the LUT with spatial independent convolution. 
%%
%Thus escaping the constraints of traversing spatial pixel combinations?

In this paper, we propose a reconstructed convolution (RC) method to decouple the spatial and channel calculations to effectively increase the RF of LUT with less storage. This decoupled operation helps network escape the constraints of traversing all spatial combinations of pixels. Thus the method can use $n\times n$ 1D LUTs to approximate the effect of a $n\times n$ convolution layer, reducing the LUT size from the original $b^{n^2}$ to $b\times{n^2}$. Based on our RC method, we present a practical Reconstructed Convolution module based LUT method (RCLUT) for SR task, enlarging RF with small storage consumption. As shown in Figure \ref{fig:PSNR&Volume}, our RCLUT achieves competitive performance and slight LUT size. It achieves a better trade-off of performance and LUT storage. Moreover, our RC method can be designed as a Plugin module in a cheap way. Although RC module sacrifices interactive information between two-dimensional features, its lightweight storage and large receptive field can well compensate for the shortcomings of previous LUT-based methods. We add RC module on top of SRLUT, which only cost about 1/10,000 of the original storage for 13 times the RF size improvement.

To conclude, the contribution of our work includes:
\begin{itemize}
    \item We propose a novel Reconstructed Convolution (RC) method with large RF, which decouples the spatial and channel calculation of convolution. Based on the RC method, our RCLUT model achieves significant performance with less storage.
    \item Our RC method can be designed as a plugin module, which brings improvement to LUT-based SR methods with slight increasing size.
    \item Extensive results show that our method obtains superior performance compared with SR methods based on LUT. It is a new state-of-the-arts LUT method in SR task.
\end{itemize}

\section{Related Work}
\label{sec:related}
\subsection{Traditional Super-Resolution}
% % Traditional single image super resolution includes 
Interpolation based methods (\eg, Nearest, Bilinear, Bicubic~\cite{bicubic}) are really efficient but tend to generate somewhat blurry results since they neglect the image content. Sparse coding based methods ~\cite{ANR, A+, image, Amultilayer, bao2013fast, gu2015convolutional} restore LR image to HR image through a learned sparse dictionary, showing promising results than interpolation based methods with a heavy run-time cost.
RAISR~\cite{raisr} learned a set of filers for different attributes of patches. This content-aware method achieves a satisfying visual result, but it has to calculate gradient and angle of patches in the prediction phase, which is not fast as interpolation methods.
\begin{figure*}[ht]
\begin{center}
% \fbox{\rule{0pt}{2in} \rule{0.9\linewidth}{0pt}}
   \includegraphics[width=17.5cm]{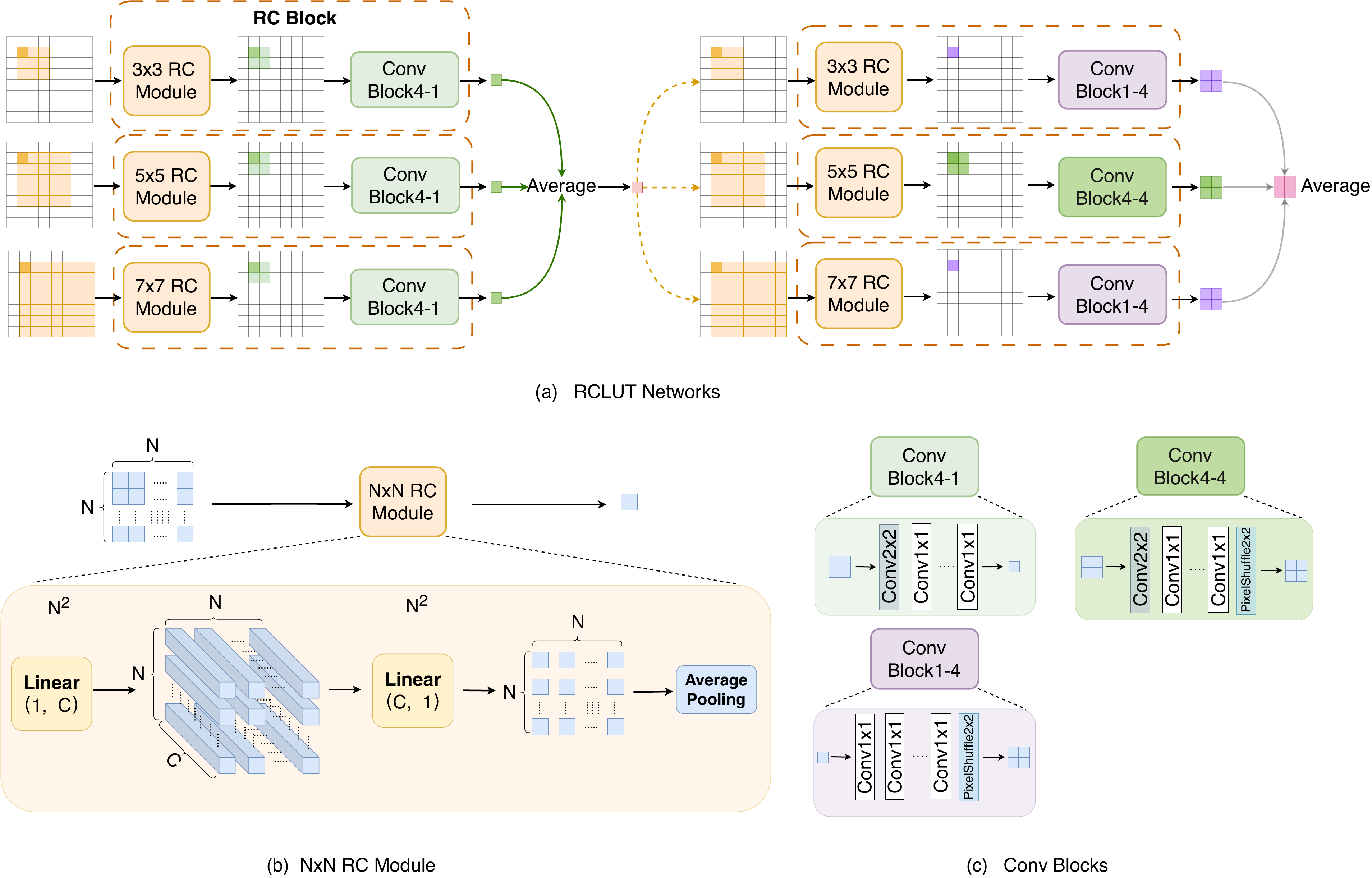}
\end{center}
   \caption{Overview of RCLUT networks. (a) The overall framework of our RCLUT networks. It contains two cascaded stages, each stage module has three branches with different receptive field RC Blocks. (b) The details of RC method for $N\times N$ RF size. It is composed of Linear and Average Pooling operations. (c) The illustration of various convolution blocks in RC Block. The input and out values of Conv Block4-1 are $4$ and $1$, respectively. The Conv Block1-4 contains $1$ input value and $4$ output value for less storage. The Conv Block4-4, which references $4$ values as input and predicts $4$ output values, is only used once in the second stage module of RCLUT.}
\label{fig:overview}
\end{figure*}
\subsection{Deep Learning Based Super-Resolution}
Deep learning based SR methods have made tremendous progress in the past decade. Starting with SRCNN ~\cite{srcnn14}, researchers propose all kinds of network architectures ~\cite{VDSR, EDSR, RCAN} to improve the performance of their models. However, these methods require huge amount of computational resources. ESPCN ~\cite{ESPCN} and FSRCNN ~\cite{FSRCNN} propose to use small networks to cut down computational cost. Others perform parameter reduction ~\cite{ahn2018fast, esrgan, chen2019camera, zhang2019two, xiao2020space, cheng2021light} or quantization ~\cite{li2020pams, song2021addersr, zhang2021edge, ayazoglu2021extremely} on state-of-the-art large SR models, with contained performance drops. From aspect of computational efficiency, AdderSR~~\cite{song2021addersr} uses addition to calculate the output features, thus avoiding the large energy consumption of traditional multiplication operations. While Namhyuk et al.~~\cite{ahn2018fast} design a lightweight and cascading residual network with a smaller number of parameters. It achieves a good trade-off in efficiency and accuracy. Nevertheless, for equipment with limited computational resources, such as mobile devices, these models are too heavy to be deployed.

\subsection{LUT Based Super-Resolution}
As mentioned in Section \ref{sec:intro}, Jo and Kim propose a pioneering work (SR-LUT), which utilizes LUTs in SR ~\cite{srlut21}. They train a simple deep SR network with limited receptive field (RF), and transfer inputs (as indices) and network outputs (positions as indices and pixel intensities as values) to a LUT. The LUT is used to generate final results during the test phase. Since there are no additional calculations required, SR-LUT can run as fast as interpolation based SR methods, \ie tens of milliseconds, on mobile devices.
Since the size of a LUT increases exponentially with the indexing capacity, the RF is limited to $3\times3$ in SR-LUT. However, the size of RF is proved crucial according to ~\cite{DBLP:conf/cvpr/GuD21} and SR-LUT naturally obtains inferior performance.
Li \etal ~\cite{mulut22} propose MuLUT, which increases the RF by introducing hand-crafted indexing patterns and cascading LUTs. By these two schemes, MuLUT achieves significant improvement over SR-LUT with the cost of a linear growth of the total LUT size.
Similarly, Ma \etal ~\cite{splut22} also adapt the idea of cascading LUTs to enlarge the RF. They propose a framework of series-parallel LUT (SPLUT) which introduces channel-level LUTs and parallelly processes the two components separated from the original 8-bits input. 
These two method devides the single LUT of SR-LUT to multi-LUTs to increase the RF of SR models.
However, their good performance is based on the linear growth of lut size at the expense of large amounts of storage resources. The trade-off between the RF and LUT size is still a challenging problem. Our method overcomes the burden of heavy LUT size when increasing the RF, thanks to the novel reconstructed convolution (RC) method, which approximates a vanilla convolution performance with exaggeratedly small LUT size. 

% figure 2 presents the architecture of RCLUT

\section{Method}
\subsection{Overview}
%LUT based SR methods have a certain limitations. As mentioned earlier, the total size of used LUTs grows fast (linearly or exponentially) with the increase of indexing capacity. This significantly restrains the RF of learned models: SR-LUT has a $3\times3$ RF; SPLUT has a $7\times7$ RF; and that of MuLUT is $9\times9$. For a simple comparison, RFs of FSRCNN ~\cite{FSRCNN} and AlexNet v2 ~\cite{AlexNet} are $195\times195$ ~\cite{araujo2019computing} and $17\times17$, respectively. As discussed in ~\cite{DBLP:conf/cvpr/GuD21}, how to effectively utilize a wide range of information is crucial for SR. These all indicate that the small RF limits the performance of LUT based SR methods.

To overcome the LUT growth of enlarging RF, we design a reconstructed convolution method. It can present $n\times n$ RF convolution into $n\times n$ 1D LUTs and have a approximate performance of vanilla convolution. 
%As mentioned in ~\cite{srlut21}, $n\times n$ RF model should be formulate $n^2$D LUTs, which storage volume is much bigger than ours.
Based on the efficient RC method, we design a Reconstructed Convolution module based LUT method, termed as RCLUT, which has parallel and cascaded structures. Besides, the RC module can be used as a Plugin module. It is flexible to improve performance of other LUT-based methods.

\subsection{Reconstructed Convolution Module}
%介绍RC的结构，实现方法，优势
Firstly, we analyse the LUT size issue of increasing RF. As discussed in ~\cite{mulut22}, the size of LUT grows exponentially when the indexing inputs increase, as $n \times n$ RF size model should be cached into $n\times n$ D LUT. In details, To convolve a 8bit $n\times n$ input to a 8bit $r\times r$ output, the full SR-LUT needs $(2^8)^{(n^2)}\times r^2$ bytes. Each pixel has $2^8$ possible values. Thus the whole $n\times n$ input has $(2^8)^{(n^2)}$ different possible combinations of pixel values. For each input, there are $r^2$ output pixels. We can see that under this formation, the size of the input becomes the key reason why the size of SR-LUT explodes. 

\begin{figure}[t]
\begin{center}
    \includegraphics[width=1\linewidth]{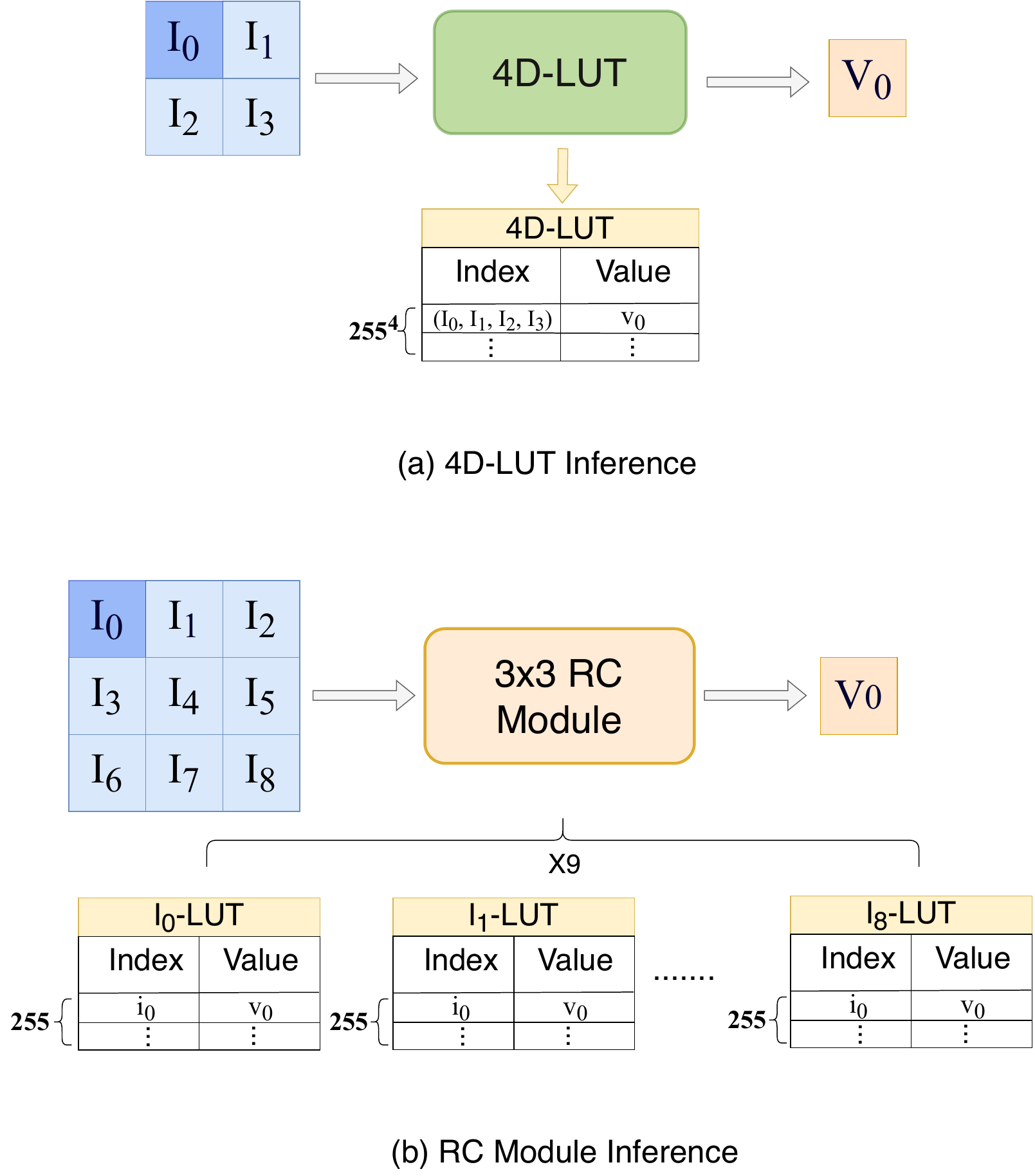}
\end{center}
   \caption{The different LUT inference formulations of SR-LUT and RC module}
\label{fig:1dlut_inference}
\end{figure}

To overcome above problem, we propose a new formulation of convolution. Vanilla convolution~~\cite{krizhevsky2017imagenet, simonyan2014very} with of $n \times n \times c\_out$ filters will calculate $n \times n$ spatial features then add the $c\_out$ results in channel dimensions. Now, we exchange the order of calculation of channel-wise features and spatial features to reconstruct convolution. It breaks the interactions between different pixels and use a small network to perform channel-wise increment and reduction on each single pixel. After that, a average pooling operation is added to generate the final feature~~\cite{chollet2017xception}. Figure~\ref{fig:overview}(b) shows the pipeline of our reconstruct method. To present a $N \times N$ RF of Reconstructed Convolution, We frist use $N^2$ Linear(c\_in=$1$, c\_out=$C$) operations to enlarge dimension of the $N\times N$ patches to obtain a feature map of $N \times N \times C$. Then, another $N^2$ Linear(c\_in=C, c\_out=1) operations reduce the channel of feature map to 1, as a $N\times N \times 1$ feature map. After the channel-wise features matching, a simple Average Pooling operation merges spatial features together to get the output value. The training process of reconstructed convolution can be formulated as
\begin{equation} \label{RC-method}
\begin{aligned}
    % \mathcal L_{m}^{s} &= \mathcal L^{cls}_m(F_{main}(E(I_m^s)), Y^s) + \mathcal L^{reg}_m(R(E(I_m^s)), Y^s)\\
    z_{(m+i, m+j)} &= W'^T_{ij}(W^T_{ij} x_{(m+i, n+j)} + b_{ij}) + b'_{ij} ,\\
    y_{mn} &= \frac{1}{N^2}\sum^{N-1}_{i=0}\sum^{N-1}_{j=0} z_{(m+i, m+j)}.
\end{aligned}
\end{equation}
Among them, $m, n$ represent the coordinates of pixels to be computed in the current feature map. $N$ is the kernel size of reconstructed convolution. $W \in R^{1\times C}$, which maps single-channel pixels into high-dimensional features. While  $W' \in R^{C\times 1}$ maps high-dimensional features back to single-channel features. $b, b'$ represent the bias. In this setting, the reconstructed convolution for each pixel can be transformed to a 1D LUT and use a Average Pooling to expand RF in a cheap way, as shown in Figure~\ref{fig:1dlut_inference}(b). Compared with 4D LUT format of SR-LUT, shown in Figure~\ref{fig:1dlut_inference}(a), our 1D LUTs have advantage of storage and inference speed. As to the inference stage, for anchor $I_0$ in Figure~\ref{fig:1dlut_inference}(b), the corresponding SR values $\boldsymbol{\mathrm{V}'}$ from reconstructed  convolution are obtained by
\begin{equation} \label{RC-index}
\begin{aligned}
 \boldsymbol{\mathrm{V}'}=\frac{1}{N^2}\sum^{N^2-1}_{i=0}(LUT_i[I_i]).
\end{aligned}
\end{equation}
%%% 冠都补充RC method的公式
While $N$ represents the size of reconstructed convolution. And it needs $N^2$ LUTs for pixel retrieval. The total size of all 1D-LUT is $2^8\times n^2\times r^2$. As shown in Table\ref{tab:size}, by transforming $n\times n$ convolutions to $n^2$ 1D LUTs (followed by global pooling), the explosion of LUT sizes with respect to RF can be successfully contained.

%%%%%%%%%%%%%%%%% method by du %%%%%%%%%%%%%
% 事实上，因为RC module将space和channel feature分开处理，缺少space和channel之间的信息交互。这对于模型的表现有一定的损害。因此我们考虑在RC module后面添加Conv Block来补充space和channel的交互信息
% RC module 自由扩大感受野且不引入大量计算负担的特性促使我们使用不同感受野大小的分支捕获更多的图像细节。
\subsection{RC-LUT Network Architecture}
RC module processes spatial and channel-wise feature separately, which lacks interactive information between both of them. As it somewhat prevents model performance, we add Conv block~\cite{DenseNet} behind the RC module to complement interactive information from the space and channel features. Specifically, we build the Reconstructed Convolution Block, which contains one RC Modue and one Conv Block~\cite{DenseNet}.  Its corresponding SR values $\boldsymbol{\mathrm{V}}$ can be obtained from $\boldsymbol{\mathrm{V'}}$,
\begin{equation} \label{lut-index}
\begin{aligned}
 \boldsymbol{\mathrm{V}_{(0,0)}}=LUT[\boldsymbol{\mathrm{V'}_{(0,0)}}][\boldsymbol{\mathrm{V'}_{(0,1)}}][\boldsymbol{\mathrm{V'}_{(1,0)}}][\boldsymbol{\mathrm{V'}_{(1,1)}}].
\end{aligned}
\end{equation}
Meanwhile, RC module has a characteristic of freely expanding the receptive field (RF) without introducing a large amount of computational burden. This prompted us to use branches of different RF sizes to capture more image details. Particularly,
we use different RF size of RC Blocks to propose a Reconstructed Convolution based LUT(RCLUT) network. The overview of our RCLUT and RC Blocks is shown in Figure~\ref{fig:overview}(a). In detals, our RCLUT network architecture has two cascaded stages. In each stages, there has three parallel branches. 
To achieve different scale feature maps like ~\cite{inception}, these RFs of three branches respectively are $3 \times 3$, $5 \times 5$, $7 \times 7$ through our RC Blocks. Moreover, in the first stage, 3 Conv Block4-1 are used after RC modules. In the second stage, one Conv Block4-4 which provides upscale ability, is quipped with RC module of $5\times5$ RF. To save total LUTs size, two Conv Block1-4 are followed by another two branches, as the Conv Block1-4 contains $(2^8)\times4$B storage as LUT format and Conv Block4-4 contains $(2^8)^4\times4$B storage without LUT-sampling.
% 冠都补充RCLUT的lut表推理形式。
With the cooperation of two cascaded stage and rotation strategy in ~\cite{mulut22,splut22}, the RCLUT network can obtain $27\times27$ RF size and maintain only $1.515$MB LUT size by sampling to $2^4$. Compared the MuLUT, The RF size of RCLUT is $9\times$ larger but the LUT size is $2.68\times$ smaller than it. Our RCLUT provides a much more efficient way to attain a trade-off of the RF and LUT size.

% figure 3 presents the architecture of RC-Plugin Module
\begin{figure*}[ht]
\begin{center}
% \fbox{\rule{0pt}{2in} \rule{0.9\linewidth}{0pt}}
   \includegraphics[width=17.5cm]{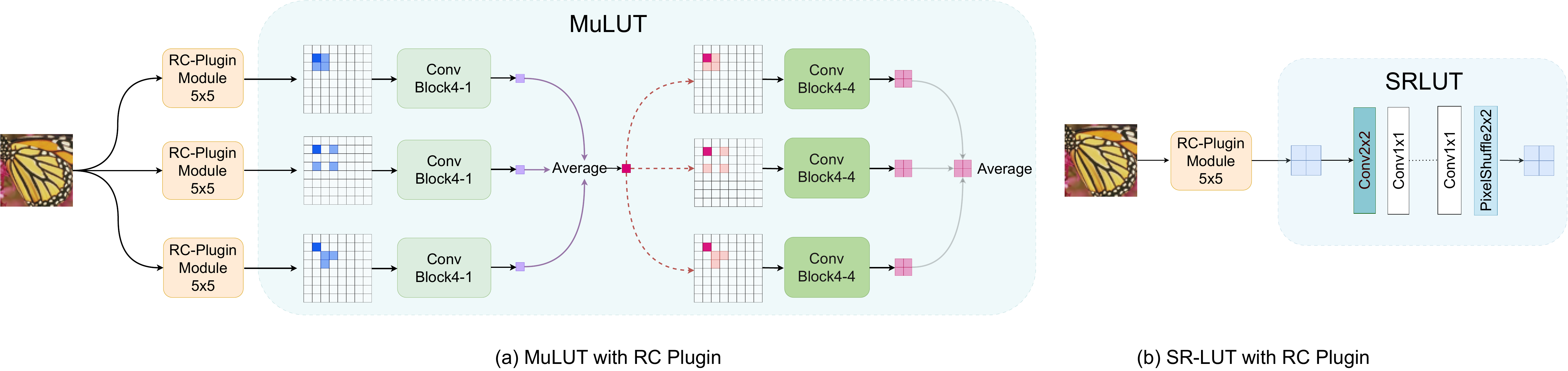}
\end{center}
   \caption{The overall framework of MuLUT and SR-LUT with RC-Plugin module. (a)As for three parallel branches architecture in MuLUT, three RC-Plugin Modules of $5\times5$ RF are put ahead of each branch. (b)SR-LUT is equipped with one RC-Plugin Module.}
\label{fig:RC_plgin}
\end{figure*}

\textbf{Discuss the receptive field (RF) size between RCLUT and MuLUT.}
% 假设RC module size为N， Conv Block接受的patch size为M，那么这两个模块耦合后的感受野大小为M+N-1，经过90旋转操作感受野大小变为了RF_1 = 2(M+N-1)-1。及联stage 2模块后，由于卷积的anchor point位置变成了中心, 经过旋转后RF_2 = 2((RF_1-1)/2 + M)-1。在RCLUT的设定下，我们选取感受野最大的分支作为网络最终的RF size。
Assuming that the RF size of RC module in 1st stage/2nd stage is $N_1/N_2$ and Conv Block is $M_1/M_2$. For RCLUT, it couples the two modules to increase the RF size to $M_1 + N_1 - 1$. Thanks to the rotation operation~~\cite{srlut21, mulut22}, the RF size of RCLUT 1st stage can further achieve to
\begin{equation} \label{rf}
\begin{aligned}
RF_1 = 2(M_1 + N_1 - 1)-1.
\end{aligned}
\end{equation}
For 2nd stage, since the anchor point in convolution process becomes the central point, $RF_1$ will shrink by half. Thus the RF size of RCLUT is
\begin{equation} \label{rf_rclut}
\begin{aligned}
RF_2 &= 2(\frac{(RF_1-1)}{2} + M_2 + N_2 - 1)-1 \\
     &= 2M_1 + 2M_2 + 2N_1 + 2N_2 - 7.
\end{aligned}
\end{equation}
While for MuLUT, its RF size increases to $2M_1-1$ through rotation. Furthermore, the cascading 2nd stage also helps to increase the RF size to
\begin{equation} \label{rf_mulut}
\begin{aligned}
RF'_2 &= 2(\frac{((2M_1-1)-1)}{2} + M_2-1)-1 \\
      &= 2M_1 + 2M_2 - 3.
\end{aligned}
\end{equation}
RCLUT's RF size can be $2N_1 + 2N_2 - 4$ more than MuULT's in the case that $M_1$ and $M_2$ of the two methods correspond equally. We select the branch where 7x7 module to calculate the final RF size of the network, where $N_1=7, M_1=2, N_2=7, M_2=1$. Under this setting, RCLUT's RF size is $27\times27$. The RF size of MuLUT is $9\times9$, where $M_1=2, M_2=2$. 

\subsection{RC-Plugin Module}
To further prove the effectiveness of the Reconstructed Convolution method, we design our RC method as a Plugin module, which has $5\times5$ RF size as shown in Figure\ref{fig:overview}(b). This RC-Plugin Module is flexible to used into LUT based SR methods(\eg SR-LUT and MuLUT). The RC-Plugin Module with $5\times5$ RF size will extend the RF size of the base LUT methods and increase only $0.415$ KB LUT size by sampling to $2^4$. To easy combine our RC-Plugin module, we only put it ahead of other LUT methods. Because of the different architectures of base models, we add three RC-Plugin modules in MuLUT and one in SR-LUT, as shown in Figure\ref{fig:RC_plgin}. It assists to enlarge MuLUT network RF size from $9\times9$ to $17\times17$.
% , enlarge SR-LUT RF size from $3\times3$ to $10\times10$.
%冠都补充SPlut w/wo plugin 感受野。

% table2. performance of sota methods 

\begin{table*}[htbp]
	\centering
        % \footnotesize
	\caption{The comparison with other methods.}
        \resizebox*{2\columnwidth}{!}{
	\begin{tabular}{ccccccccc}
		\toprule  % 顶部线
             & Method & RF Size & LUT size & Set5 & Set14 & BSDS100 & Urban100 & Manga109 \\
            \midrule
            \multirow{3}*{Interpolation} & Nearest & $1\times1$ & - & 26.25/0.7372 & 24.65/0.6529 & 25.03/0.6293 & 22.17/0.6154 & 23.45/0.7414 \\
            ~ & Bilinear & $2\times2$ & - & 27.55/0.7884 & 25.42/0.6792 & 25.54/0.6460 & 22.69/0.6346 & 24.21/0.7666 \\
            ~ & Bicubic & $4\times4$ & - & 28.42/0.8101 & 26.00/0.7023 & 25.96/0.6672 & 23.14/0.6574 & 24.91/0.7871 \\
            \midrule
            \multirow{4}*{LUT} & SR-LUT~~\cite{srlut21} & $3\times3$ & 1.274MB & 29.82/0.8478 & 27.01/0.7355 & 26.53/0.6953 & 24.02/0.6990 & 26.80/0.8380 \\
            ~ & SPLUT-L~~\cite{splut22} & & 18MB & 30.52/0.8631 & 27.54/0.7520 & 26.87/0.7091 & 24.46/0.7191 & 27.70/0.8581 \\
            ~ & MuLUT~~\cite{mulut22} & $9\times9$ & 4.062MB & 30.60/0.8653 & 27.60/0.7541 & 26.86/0.7110 & 24.46/0.7194 & 27.90/0.8633 \\
            ~ & RCLUT (Ours) & $27\times27$ & 1.513MB & \textbf{30.72/0.8677} & \textbf{27.67/0.7577}	& \textbf{26.95/0.7145} & \textbf{24.57/0.7253} & \textbf{28.05/0.8655} \\
            \midrule
            \multirow{4}*{Sparse coding} & NE + LLE~~\cite{NE+LLE} & - & - & 29.62/0.8404 & 26.82/0.7346 & 26.49/0.6970 & 23.84/0.6942 & 26.10/0.8195 \\
            ~ & Zeyde~~\cite{Zeyde} & - & - & 26.69/0.8429 & 26.90/0.7354 & 26.53/0.6968 & 23.90/0.6962 & 26.24/0.8241 \\
            ~ & ANR~~\cite{ANR} & - & - & 29.70/0.8422 & 26.86/0.7368 & 26.52/0.6992 & 23.89/0.6964 & 26.18/0.8214 \\
            ~ & A+~~\cite{A+} & - & - & 30.27/0.8602 & 27.30/0.7498 & 26.73/0.7088 & 24.33/0.7189 & 26.91/0.8480 \\
            \midrule
            DNN & FSRCNN~~\cite{FSRCNN} & $17\times17$ & - & 30.72/0.8660 & 27.61/0.7550 & 26.98/0.7150 & 24.62/0.7280 & 27.90/0.8610 \\

		\bottomrule  % 底部线
	\end{tabular}
 }
	\label{tab:psnr_ssim}
\end{table*}

\section{Experiment}
In this section, we first introduce the datasets and training details of RCLUT networks, and we compare our RCLUT to several state-of-the-art LUT based SR methods by quantitative and qualitative evaluation. Besides, the quantitative experiments will introduce the benefits of RC-Plugin module to other LUT methods. Finally, we perform ablation studies to validate the effectiveness of our RCLUT model and RC module.
\subsection{Datasets and Experimental Setting.} 
\textbf{Datasets and Metrics.}
We train the RCLUT Networks on the DIV2K dataset~\cite{DIV2K}, which contains 2K resolution images for SR task. The widely used 800 training images in DIV2K are chosen to train RCLUT model. We focus on the $\times 4$ upscale factor in SR task, which LR images are simply downscaled by \textit{Bicubic} interpolation. Public benchmarks of Set5, Set14, BSD100~\cite{Set}, Urban100~\cite{Urban100} and Manga109~\cite{Manga109} evaluate the performance of our method. To the fair comparison, only Peak Signal-to-Noise Ratio(PSNR) and structural similarity index(SSIM)~\cite{wang2004image} both at Y-channel are used as the evaluation metrics. Besides, we involve storage of LUTs to evaluate the efficiency of other methods.

\textbf{Training Setting.}
Our training setting is as follows. The channel increasing number $C$ in RC Linear Operation is set to $64$. Our RCLUT model is trained for $200000$ iterations with Adam optimizer~\cite{kingma2014adam} with learning rate of $1e^{-4}$ and a batch size of $32$. The mean-squared error (MSE) loss function is selected as the optimization target.  We also maintain the rotation training strategy to improve performance and the RF. The Data augmentation of random flip and rotation is used to enhance our model ability. We train RCLUT model with PyTorch~\cite{stevens2020deep} on Nvidia V100 GPUs.\par
\textbf{Caching LUT Setting.}
When the model is converged, we convert RCLUT networks to multi-LUTs format with interval $2^4$ to decrease the size. 
Because of the cascade strategy, results of first stage should be quantized to integers, we use the same Re-indexing method in MuLUT. Moreover, the LUT-aware Finetuning Strategy proposed by MuLUT is also used to keep LUTs performance same with networks.

\subsection{Quantitative Evaluation}
We compare our method with various SR methods, including 3 interpolation based methods(nearest neighbor, bilinear and bicubic interpolation), 4 sparse coding methods(NE+LLE~\cite{NE+LLE}, Zeyde \etal ~\cite{Zeyde}, ANR~\cite{ANR} and A+~\cite{A+}), 1 DNN-based method(FSRCNN~\cite{FSRCNN}) and 3 LUT-based method(SR-LUT~\cite{srlut21}, MuLUT~\cite{mulut22} and SPLUT~\cite{splut22}).

The quantitative results are shown in Table~\ref{tab:psnr_ssim}. As observed, with the assistance of much larger RF size, our RCLUT model achieves the highest PSNR performance in LUT-based methods and sparse coding methods. Compared with prior state-of-the-art method~\cite{mulut22}, we improve 0.12dB PSNR value on Set5 dataset with the benefit of $27\times 27$ RF size. Moreover, RCLUT outperforms FSRCNN on Set14 and Managa109 benchmark dataset, and obtains more comparable performance in the other datasets.

To compare with the efficiency, we also report the statistics of multiplications and additions and LUT storage size of LUT-based methods. As shown in Table~\ref{tab:psnr_ssim}, our RCLUT shows progressive performance but has similar LUT size with SR-LUT, which is $0.39$ MB larger than SR-LUT. Besides, RCLUT LUT size is $11.88 \times$ smaller than SPLUT-L but achieves $0.2$dB higher PSNR value in Set5 dataset. Though the performance is improved by 0.11dB PSNR value on average, RCLUT size is only $2.68\times$ smaller than MuLUT.

The extensive comparisons of performance and LUT size verify that our RCLUT is more efficient method than previous LUT-based methods. It provides a better trade-off between accuracy and storage. The RCLUT references much more pixels information to ensure the restoration ability with less increasing size via the RC module.

% figure 4 illustrates the qualitative evaluation
\begin{figure*}[t]
\begin{center}
% \fbox{\rule{0pt}{2in} \rule{0.9\linewidth}{0pt}}
   \includegraphics[width=17cm]{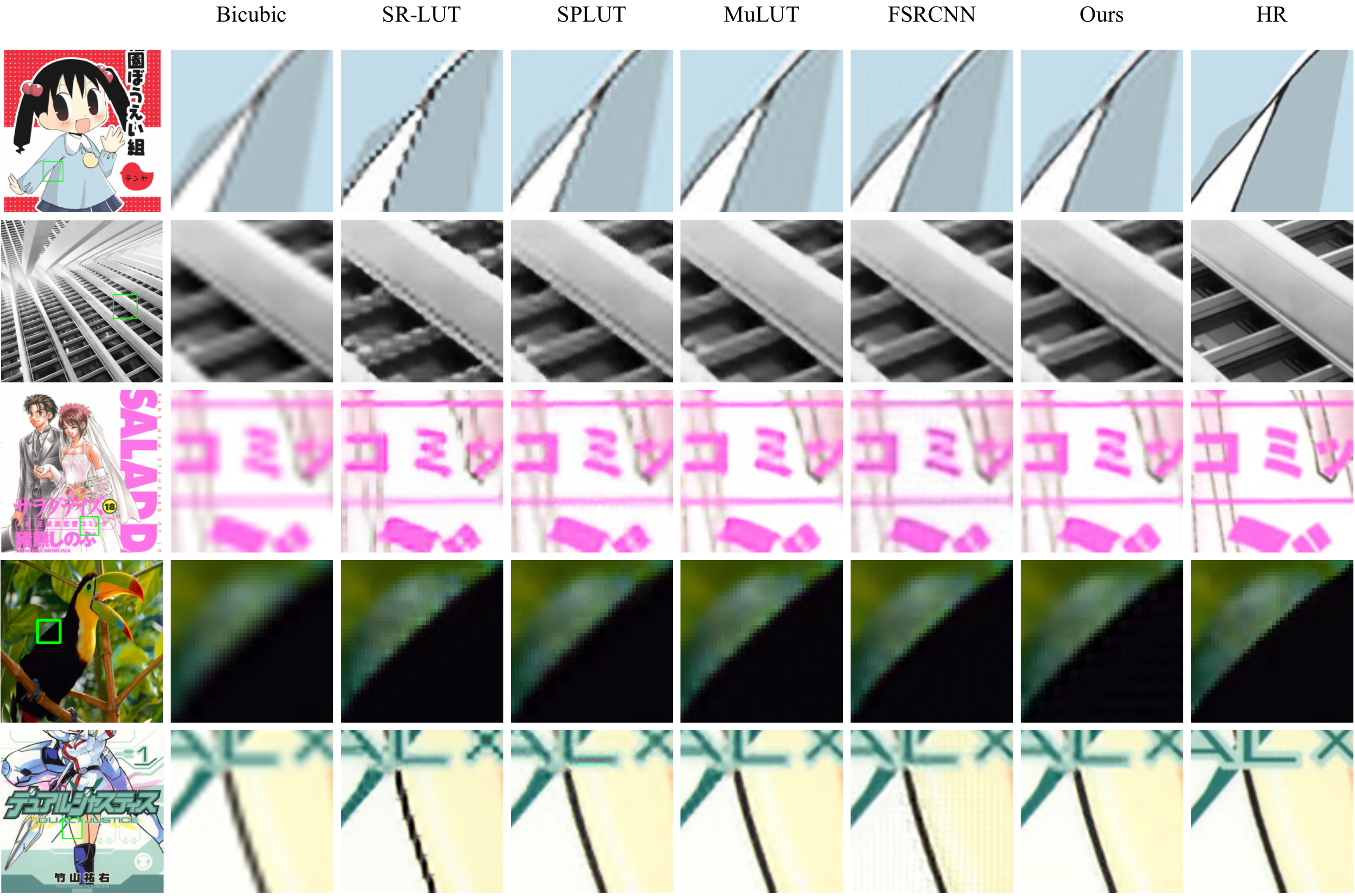}
\end{center}
   \caption{Visual comparison for $\times 4$ SR task on benchmark datasets. The results show our RCLUT can generate sharp edges without severe artifacts compared with other methods.}
\label{fig:visual_examples}   
\end{figure*}
\begin{table*}[h!]
	\centering
        % \footnotesize
	\caption{SR-LUT and 1D-LUT size estimation when storing 8bit output for 8bit input with upscaling factor $r=4$. SR-LUT becomes unpractical when RF is larger than $2\times 2$.}
	\begin{tabular}{cccc}
		\toprule  % 顶部线
            RF & Full size SR-LUT & Sampled SR-LUT & Full size 1D-LUT \\
            \midrule
            $2\times2$ & 64GB & 1.274MB & 16KB \\
            $3\times3$ & $6.7\times10^7$PB & 1.726TB & 36KB \\
            $5\times5$ & $2.3\times10^{46}$PB & $8.2\times10^{16}$PB & 100KB \\
            $n \times n$ & $(2^8)^{(n^2)}\times r^2$B & $(2^4+1)^{(n^2)}\times r^2$B & $2^8\times n^2\times r^2$B \\
		\bottomrule  % 底部线
	\end{tabular}
	\label{tab:size}
\end{table*}
\begin{table*}[h]
\centering
\caption{The effectiveness RC-Plugin module.}
\resizebox*{1.5\columnwidth}{!}{
\begin{tabular}{c|c|c|c|c|c|c}
\toprule
Method & Set5 & Set14 & BSDS100 & Urban100 & Manga109 & Volume\\
\midrule
MuLUT~~\cite{mulut22} & 30.60 & 27.60 & 26.86 & 24.46 & 27.90 & 4.062 MB\\
RC-5 + MuLUT & \textbf{30.77} & \textbf{27.71} & \textbf{26.96} & \textbf{24.58} & \textbf{28.11} & 4.062 MB (+ 1.245 KB)\\
\bottomrule
\end{tabular}
}
\label{tab:mulut_plgin}
\end{table*}

\subsection{Qualitative Evaluation}
We mainly compare our method with 3 LUT-based SR methods and FSRCNN network in this section. Figure~\ref{fig:visual_examples} illustrates visual quality of 5 cases in benchmark datasets, and more visual results are provided in supplementary material. As shown in the first and the last example, Bicubic interpolation brings blurry results, SR-LUT involves obvious blocking artifacts because of the limited RF size, FSRCNN brings serious checkerboard artifacts, MuLUT and SPLUT present relative better quality than SR-LUT but still with noises. On the constrary, our methods achieves more satisfactory results than others, it illustrates sharper edges and less artifacts. The other three examples also show that, our RCLUT model recovers clearer edges and more natural textures even compared with DNN method (FSRCNN). These progressive visual results verify the effectiveness of RCLUT with exaggerate RF size.

\subsection{The effectiveness of RC-Plugin Module}
% 和sr-lut结合，和mu-lut结合
One more experiment verities RC method can be used as a plugin module, it is efficient and flexible to benefit to other LUT-based methods. We use $5\times5$ RF size RC module as RC-Plugin module, termed as RC-5, and put one RC-Plugin module ahead of SR-LUT, and put three RC-Plugin module ahead of each branch in MuLUT, as illustrated in Figure~\ref{fig:RC_plgin}. Table \ref{tab:mulut_plgin} and the first and the third row in Table \ref{tab:size} show that, equipped with RC-Plugin module, RC-5+SR-LUT and RC-5+MuLUT gain respectively $0.54$dB and $0.17$dB PSNR value over the base methods on Set5 dataset. As the table size in Table~\ref{tab:mulut_plgin} and Table~\ref{tab:size} shows, the LUT size is negligibly increased after equipped with RC-Plugin module. It verities that our RC-Plugin module is effective and efficient.

\subsection{Ablation Studies}

In this section, we discuss the effectiveness of RCLUT networks and the implement of RC module.
\begin{table*}[!ht]
\centering
\caption{Ablation study of RC module.}
\resizebox*{1.9\columnwidth}{!}{
\begin{tabular}{c |c|c|c|c|c|c|c|c|c}
\toprule
Avg Pooling 5x5 & RC-5 & Conv Block4-4 & Conv Block1-4 & Set5 & Set14 & BSDS100 & Urban100 & Manga109 & Volume\\

\midrule
 & & \Checkmark & & 29.82 & 27.01 & 26.53 & 24.02 & 26.80 & 1.274 MB\\
\Checkmark & & \Checkmark & & 24.19	& 23.09	& 23.71	& 20.78	& 21.88 & 1.274 MB \\
 & \Checkmark & & \Checkmark & 29.49	& 26.80 & 26.44 & 23.77 & 26.09 & 0.83 KB \\
 & \Checkmark & \Checkmark & & \textbf{30.40} & \textbf{27.40} & \textbf{26.77}	& \textbf{24.26} & \textbf{27.33} & 1.274 MB \\
\bottomrule
\end{tabular}
\label{tab:RC_method}
}
\end{table*}

\textbf{The effectiveness of Linear operations in RC module.}
Our Reconstructed Convolution is based on Linear and Average Pooling operations. This section will show the influence of Linear operations in RC module. 
As the Table~\ref{tab:RC_method} shows, followed by Conv Block4-4, the method only equiped with Average Pooling can also get the $5\times5$ RF size, but it decreases the ability of Conv Block4-4(SR-LUT baseline). On the contrast, the $5\times5$ RC module with Linear and Average Pooling operations, termed as RC-5, improves the behind method over $6.21$dB of PSNR value on Set5 dataset. This experiment conducts the channel-wise feature is import to the SR model, our Linear operations provides a matching ability to obtain depth features. On the other hand, the combination of Conv Block1-4 and RC-5 contains performance drops than model of Conv Block4-4 and RC-5. We analyse that our RC module is lack of spatial features fusion, it needs to cooperate with vanilla convolution blocks, like Conv Block4-4, to have complement effect. 

\begin{table}[!ht]
\centering
\caption{Ablation study of receive field size on RC.}
\resizebox*{1\columnwidth}{!}{
\begin{tabular}{c |c|c|c|c|c|c}
\toprule
Method & Set5 & Set14 & BSDS100 & Urban100 & Manga109 & Volume\\

\midrule
SR-LUT~~\cite{srlut21} & 29.82 & 27.01 & 26.53 & 24.02 & 26.80 & 1.274 MB \\
RC-3 + SR-LUT & 30.34 & 27.37 & 26.75 & 24.24 & 27.28 & 1.274 MB(+0.149 KB)\\
RC-5 + SR-LUT & 30.40 & 27.40 & 26.77 & 24.26 & 27.33 & 1.274 MB(+0.415 KB)\\
RC-7 + SR-LUT & 30.43 & 27.44 & 26.80 & 24.31 & 27.44 & 1.274 MB(+0.813 KB)\\
RC-9 + SR-LUT & 30.45 & 27.46 & 26.81 & 24.33 & 27.48 & 1.274 MB(+1.345 KB)\\
\bottomrule
\end{tabular}
}
\label{tab:rf_RCLUT}
\end{table}

\textbf{The effectiveness of RF of RC module.}
We test the influence of the RF size from $3\times3$ to $9\times9$ to RCLUT model. These experiments only test on the one stage and one branch network architecture for easy implement. Table~\ref{tab:rf_RCLUT} shows that, with benefit of $3\times3$ RC module, termed as RC-3, SR-LUT increases $0.48$ dB of PSNR value on Set5 dataset. At the same time, $5\times5$ RF size and $7\times7$ RF size can improve PSNR value by $0.06$dB and $0.09$dB over RC-3 module, respectively. On the other hand, $9\time9$ RF size of RC module only increase $0.02$ PSNR value over RC-7. These results show that, with the increasing RF, our RC module brings improving performance, and the inflection point of RF is near to $7\times7$. It is reason why we choose the maximum of RF size in RC Blocks as $7\times7$. 

\begin{table}[!htbp]
\centering
\caption{Ablation study of multi-branch on RC.}
\resizebox*{1\columnwidth}{!}{
\begin{tabular}{c |c|c|c|c|c|c}
\toprule
Method & Set5 & Set14 & BSDS100 & Urban100 & Manga109 & Volume \\
\midrule
RCLUT-3 & 30.34	& 27.37 & 26.75 & 24.24 & 27.28 & 1.274 MB \\
RCLUT-3\_5 & 30.57 & 27.57 & 26.89 & 24.46 & 27.68 & 2.548 MB \\
RCLUT-5\_7 & 30.64	& 27.63 & 26.93 & 24.53 & 27.85 & 2.548 MB \\
RCLUT-3\_5\_7 & 30.68 & 27.64 & 26.94 & 24.55 & 27.90 & 3.822 MB \\
RCLUT-5\_7\_9 & 30.68 & 27.63 & 26.94 & 24.56 & 27.93 & 3.822 MB \\
RCLUT-3\_5\_7-X2 & \textbf{30.72} & \textbf{27.67} & \textbf{26.95} & \textbf{24.57} & \textbf{28.05} & 1.513 MB \\
\bottomrule
\end{tabular}
}
\label{tab:branches_RCLUT}
\end{table}

\textbf{The effectiveness of cascade and parallel branches in RCLUT.}
We take RCLUT-3 as single stage model with $3\times3$ RC module and Conv Block4-4, RCLUT-3\_5 as single stage model with two branches of $3\times3$ and $5\times5$ RC Blocks. In Table~\ref{tab:branches_RCLUT}, We can observe that multi-branch module improves the performance of RCLUT, especially RCLUT-3\_5\_7 obtains 0.34dB higher PSNR value over RCLUT-3 on Set5 dataset. Another interesting phenomenon is that RCLUT-5\_7\_9 has similar performance with RCLUT-3\_5\_7, we believe that the  $7\times7$ RF size is enough for our RCLUT, it is the same with above section. Moreover, the two cascaded stages model RCLUT-3\_5\_7-$X\times2$ improves 0.04dB of PSNR performance on Set5 dataset than RCLUT-5\_7\_9, that of single stage model. These experiments explain the effectiveness of cascade and parallel branches architecture of RCLUT.

\begin{table}[h!]  
\small
\centering
\vskip -0.3cm
\resizebox{1\columnwidth}{!}{
\begin{tabular}{c |c|c|c|c|c|c}
    \toprule
     & \multicolumn{5}{|c|}{LUT} & DNN \\
    \midrule
     ~ & SR-LUT & SR-LUT$^{\Omega}$ & MuLUT & MuLUT$^{\Omega}$ & RC-LUT & EDSR \\
    \midrule
     Runtime (ms) & $152$ & $157(+3\%)$ & $253$ & $266(+5\%)$ & $232$ & $8083$ \\
    \hline
\end{tabular}
}
\label{table:speed}
\vskip -0.4cm
\end{table}
\textbf{The inference speed of LUT-based methods.} we conduct experiments following the same setup as MuLUT [24] by generating a 1280 × 720 HD image through $4\times$ SR and present the comparison. ${\Omega}$ represents the RC plug-in module with $5\times5$ receptive field. 
EDSR is implemented in the CPU-version of the PyTorch library and its runtime is measured on a laptop (MacBook Pro with 2.6 GHz Intel Core i7). All the other runtimes are measured on an IQOO Neo5 smartphone. It should be noted that generally PC processors are more powerful than mobile processors. 
\section{Conclusion}
In this paper, we propose a Reconstructed Convolution method, which decouples the calculation of spatial and channel-wise feature. The RC method can be formulated as multi-1D LUTs to maintain large RF with less LUT size. Moreover, We propose a RCLUT model based on RC Blocks. It achieves significant performance with slight increasing LUT size. On the other hand, our RC method can be used as a Plugin module to improve the ability of LUT-based SR methods. Extensive experiments show that our RCLUT is a new the-state-of-art LUT-method for SR task, it works much more efficiently than other prior methods.

{\small

\bibliographystyle{ieee_fullname}
\bibliography{egbib}
}

\end{document}